# CRAFT: A Multifunction Online Platform for Speech Prosody Visualisation

Dafydd Gibbon

Universität Bielefeld, Germany
gibbon@uni-bielefeld.de

**ABSTRACT**

There are many research tools which are also used for teaching the acoustic phonetics of speech rhythm and speech melody. But they were not purpose-designed for teaching-learning situations, and some have a steep learning curve. *CRAFT* (*Creation and Recovery of Amplitude and Frequency Tracks*) is custom-designed as a novel flexible online tool for visualisation and critical comparison of functions and transforms, with implementations of the *Reaper*, *RAPT*, *PyRapt*, *YAAPT*, *YIN* and *PySWIPE* F0 estimators, three *Praat* configurations, and two purpose-built estimators, *PyAMDF*, *S0FT*. Visualisations of amplitude and frequency envelope spectra, spectral edge detection of rhythm zones, and a parametrised spectrogram are included. A selection of audio clips from tone and intonation languages is provided for demonstration purposes. The main advantages of online tools are consistency (users have the same version and the same data selection), interoperability over different platforms, and ease of maintenance. The code is available on GitHub.

**Keywords**: tone, intonation, rhythm zone, f0 estimation, pitch extraction, prosody visualisation

## 1. INTRODUCTION

There are several excellent tools which are often used not only for research but also for teaching the acoustic phonetics of speech rhythm and speech melody. The most popular and versatile is *Praat* [4]. An advantage of *Praat* is the conceptually clear, but, for most new users, unfamiliar *object+methods* user interface. Other tools such as *WinPitch* [25] and *Annotation Pro* [23] have familiar user interfaces but are platform-restricted, or are strictly purposed, such as *ProsodyPro* [41], while tools such as *WaveSurfer* [32] are dedicated research environments. These tools are offline applications, with the advantage of efficiency and the disadvantage of version plurality.

The online and platform-independent *CRAFT* (*Creation and Recovery of Amplitude and Frequency Tracks*) visualisation tool was developed to overcome some of the drawbacks noted above, and to focus on teaching-learning environments in linguistic and acoustic phonetic pedagogy (as opposed to pronunciation teaching). *CRAFT* is solely a parametrised visualisation tool, not a full phonetic workbench. The pedagogical motivation for developing *CRAFT* as an easy-to-use and interoperable online tool stems from 'how does it work' questions in advanced acoustic phonetics classes, from an explicit requirement to instil a critical and informed initial understanding of the strengths and weaknesses of different speech signal visualisations, and from a practical need for an easily accessible tool for distance tutoring and face-to-face teaching, usable on laptops, tablets and smartphones.

The specifications of *CRAFT* are described in Section 2, with use cases of F0 estimation, envelope spectral analysis ('rhythm' modelling) and algorithm evaluation in Sections 3, 4 and 5, and a summary and an outlook outline in Section 6.

## 2. SPECIFICATION

### 2.1. Use cases

The main visualisations provided by *CRAFT* are:
1. speech melody: selected F0 estimation ('pitch' tracking) algorithms, with a parametrisation option,
2. rhythm: amplitude and frequency envelope spectra.

The primary criteria for the choice of an online teaching tool are, assuming browser interoperability:
1. there is no version inconsistency at any given time,
2. users are not restricted to specific computer types,
3. interoperability of the software extends to tablets and (with screen size limitations) mobile phones,
4. ubiquity in distributed distance learning is given,
5. protocols of algorithm performance can easily be created and collated,
6. maintenance is facilitated by server-side operation.

### 2.2. Architecture (GUI and system)

Interaction with *CRAFT* is via a standard HTML input form, divided into five panels (cf. The online demonstration[1] and Figure 1):
1. the main input frame, with parameters for signal processing, in particular F0 estimation,
2. three frames for amplitude and frequency modulation, demodulation and comparison of performance of F0 extractors and components, illustrations of filter types, Fourier and Hilbert Transforms, and a parametrised spectrogram,
3. the output display frame.

---

1  *http://wwwhomes.uni-bielefeld.de/gibbon/CRAFT/*

**Figure 1:** *CRAFT* graphical user interface: parameter input frame (top) showing 9 F0 extractor algorithms; amplitude demodulation (left upper mid); F0 estimator comparison (left lower mid); filter, transformation, parametrised spectrogram (left bottom); output frame (lower right).

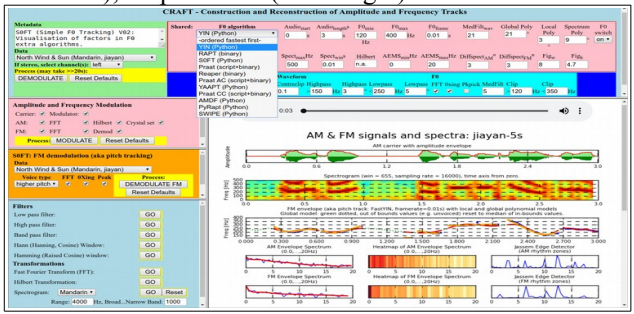

## 2.3. Implementation

*CRAFT* is designed and coded as a client-server HTML-CGI application in Python 2.7, The *S0FT*, *PyAMDF*, *FastYIN*, *PyYAAPT*, *PyRapt* and *PySWIPE* F0 estimators and most other routines are coded in Python 2.7 and interoperable on all major platforms. Exceptions are precompiled binaries for the *Reaper, RAPT* and *Praat* F0 estimators with Python wrappers, which currently run only under Linux. Python has a relatively shallow learning curve, allowing use of the code in signal processing tutorials, for which implementation criteria such as speed, size and efficiency are less important than clarity. A minimalistic functional programming style is used. *CRAFT* can easily be extended for more specialised teaching, cf. [43], or for research as described in Section 5; cf. also [13].

The online demonstration of *CRAFT* with corpus snippets of *The North Wind and the Sun* in Mandarin and English runs under Solaris on a central server. The code also runs with local web servers, so that online, offline and intranet uses are also options. The source code is freely available on *GitHub*. An offline package suitable for more advanced work is provided by Reichel's *CoPaSul* [30].

## 3. MELODY: F0 ESTIMATION

Several comparisons and analyses of F0 estimators are discussed in [2]. The CRAFT environment gives students the opportunity to perform such analyses for themselves. An optimal F0 representation is often defined by bench-marking against a gold standard [16], [29], [20] such as laryngograph output or establshed F0 estimators such as *RAPT* or *PRAAT*. In the absence of a laryngograph, implementations of nine *de facto* standard algorithms and algorithm settings were included for analysis and comparison, as well as two purpose-built algorithms (the new *CREPE* [22] neural net F0 estimator is not included):

1. *Praat* [4]: one cross-correlation and two autocorrelation configurations,
2. *RAPT*, [33], cross-correlation,
3. *PyRapt*, a Python emulation of *RAPT* [11],
4. *Reaper* (*Robust Epoch and Pitch EstimatoR*), vocal cord closure epoch detection [34],
5. *SWIPE* [10], spectrum-based, in Python [5],
6. *YAAPT* [42], hybrid frequency and time-domain methods, in the *PyYAAPT* emulation [31],
7. *YIN* [7] autocorrelation, as *FastYIN* in Python [14],
8. AMDF (*PyAMDF*), purpose-designed for *CRAFT*,
9. Purpose-designed parametrised estimator (*S0FT*).

Frequently asked questions about F0 estimation ('pitch tracking') in teaching situations concern gaps and diversions from a perceived smooth pitch trajectory. The main answer offered by *CRAFT* is to provide the *Simple F0 Tracker* (*S0FT*) using basic signal processing methods (cf. [16], [24]):

1. Preprocessing:
   1. adjustable centre-clipping (numerical censoring) to reduce low amplitude higher frequencies;
   2. low-pass filtering to reduce the magnitude of remaining higher frequency harmonics;
   3. high-pass filtering to reduce low-frequency noise.
2. F0 estimation:
   1. FFT windowing with identification of strongest and lowest harmonic peak;
   2. zero-crossing interval measurement;
   3. peak-picking interval measurement converted to zero-crossing measurement by differencing.
3. Post-processing:
   1. sample clipping outside a defined frequency range;
   2. median smoothing of the detected F0.

These eight *S0FT* parameters define a large processing space and are all adjustable by the student, with the goal of empirically tweaking values to obtain an optimal F0 representation for the data provided. Trial and error is no substitute for formal understanding, but it contributes to motivation and to the development of analytic intuitions.

A further step in F0 analysis involves abstractions over aspects of phonation and pitch perception. Fujisaki's model is production-oriented [9] while other approaches (Hirst's quadratic spline interpolation [8], the IPO model [35] or Mertens' *Prosogram* (cf. [1], [27]) are implicitly or explicitly perception oriented. Cubic polynomials have been successfully used as abstract models for the lexical tones of Thai [36] and Mandarin [40], [25]. For phrasal F0 trajectories higher order polynomials are needed. *CRAFT* therefore includes options for two polynomial modelling domains (Figure 2):

1. polynomial models of non-zero segments of utterances, to show F0 tendencies relating to tones, pitch accents and stress correlates,

2. polynomial models of the entire utterance: median F0 is calculated, ignoring zero segments, and used to interpolate over voiceless segments (yielding similar results to spline interpolation [17]).

**Figure 2:** F0 estimates by the *S0FT* and *RAPT* F0 estimators, with polynomial models (local: orange; global: red with dotted green linking lines).

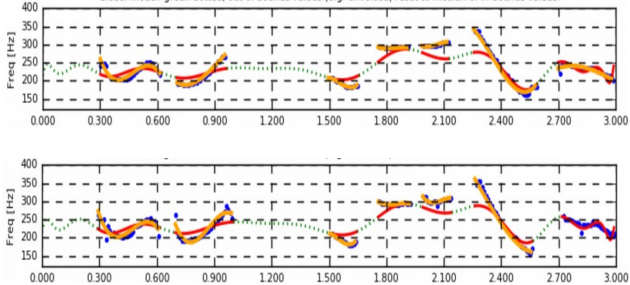

Visual inspection shows that the empirically tweaked *S0FT* F0 trajectory and polynomial models correspond closely to those derived from other algorithms. Correlations vary with different data samples (in general $0.7 < r < 0.9$ with Pearson's $r$ as a simple similarity measure; see Section 5).

## 4. RHYTHM: ENVELOPE SPECTRA

There are many approaches to rhythm description: grammatical (metrical, optimality-theoretic [21]); phonological (intonation and tone with oscillation modelled abstractly as regular iteration [28]); annotation-based isochronous interval models of irregularity, e.g. variability measures (cf. overview in [13]), or phonetic oscillator models [3], [6], [18].

Many oscillator rhythm models are production oriented. In conventional signal processing terms, they model a (laryngeal) carrier frequency and the (mainly supralaryngeal) *amplitude modulation* (AM) frequencies of syllables and phrases. *CRAFT* takes the complementary perception modelling approach of *amplitude demodulation* (cf. [15], [39], [37], [38]) by tracing the modulation envelope and applying spectral analysis by FFT to recover the AM frequencies. Formally, amplitude demodulation is the absolute Hilbert Transform of the signal, but for teaching purposes a simpler 'crystal radio set detector' procedure is used: in a parametrised moving window over the absolute (rectified) signal the maximum is selected, followed by global low-pass filtering, yielding an accurate envelope. An FFT is applied to the AM envelope in order to derive the Amplitude Envelope Spectrum (AES).

The relation of the AM modulation spectrum to speech rhythms is conceptualised in a *Multiple Rhythm Zone* model (cf. [12], [13]): different frequency segments in the spectrum represent different kinds of highly variable 'fuzzy' rhythm, e.g. of phones, of syllables, feet, phrases, interpausal units and discourse sequences, are defined in terms of overlapping rhythm zones. (Figure 3). The AES is differenced to mark boundaries between rhythm zones (*Jassem Edge Detection*, named after the pioneer of speech segment edge detection by differencing [19]).

**Figure 3:** AES, FES and rhythm zone marking with Jassem Edge Detection (x-axis in Hz).

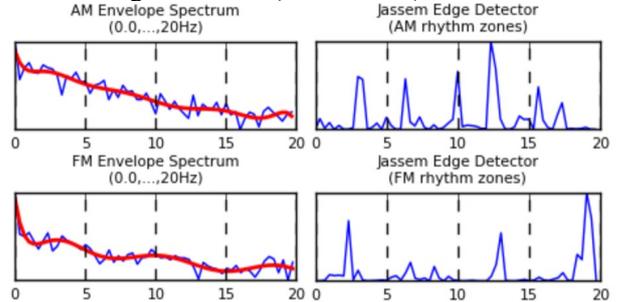

Similarly, the F0 contour is conceptualised as *frequency modulation* (FM). F0 estimation or 'pitch' tracking is conceptualised as *frequency demodulation*, the extraction of the frequency modulation envelope (FES), to which spectral analysis and edge detection are also applied.

## 5. USE CASE: ALGORITHM EVALUATION

The relative contributions of AM and FM to the production, transmission and perception of rhythm remain to be investigated. *CRAFT* is a suitable exploratory tool for this task. To illustrate this use case, AM and FM envelopes and their envelope spectra were used in an initial investigation. For the comparison of AM and FM envelopes the lengths of the vectors were normalised and the F0 vector was median-interpolated. Correlations are roughly comparable, averaging around $r = 0.7$. Global AM and FM correlations are unsurprising: syllables are largely co-extensive with tones and pitch accents.

**Table 1:** Selected F0 estimator correlations for S0FT, RAPT, PyRapt and Praat on a single data sample.

| Correlation | $r$ | $p$ |
|---|---|---|
| S0FT:RAPT | 0.897 | << 0.01 |
| S0FT:PyRapt | 0.807 | << 0.01 |
| S0FT:Praat | 0.843 | << 0.01 |
| RAPT:PyRapt | 0.883 | << 0.01 |
| RAPT:Praat | 0.868 | << 0.01 |
| PyRapt:Praat | 0.791 | << 0.01 |

The pedagogical value of algorithm evaluation, using the same data, was illustrated using *CRAFT* numerical output, again with Pearson's $r$ as the similarity measure with length normalisation and and with median interpolation. In the snapshot of a single speech sample shown in Table 1, the manually optimised *S0FT* result compares well with *RAPT*, as a gold standard, and with *Praat* and the *PyRapt* emulation of *RAPT*. Far-reaching conclusions cannot be drawn from such individual cases, of course.

Different F0 estimator implementations differ in median processing time with the same data, depending on their window length and skip sizes, and use of frequency vs. time domain techniques (snapshot with i7-7700K CPU, 100 iterations: 0.02s...14.58; cf. Table 2). Interestingly, the fastest, *FastYIN*, in Python, is faster than the *RAPT* binary and is evidently highly optimised, using Python libraries implemented in Fortran. The three slowest are implemented in Python with standard techniques.

Table 2: F0 modules ordered by processing times.

| Name | Method (simplified) | $T_{proc}$ (s) |
|---|---|---|
| *FastYIN* | Autocorrelation | 0.0181 |
| *RAPT* | Normalised crosscorrelation | 0.0217 |
| *S0FT* | Zero crossings ✕ FFT peaks | 0.0544 |
| *Praat* | Autocorrelation (general default) | 0.1087 |
| *PraatAC* | Autocorrelation | 0.1311 |
| *Reaper* | Normalised crosscorrelation | 0.1523 |
| *PyYAAPT* | Modified crosscorr, FFT peaks | 0.2394 |
| *PyAMDF* | Average Magnitude Diff Function | 0.2705 |
| *PraatCC* | Crosscorrelation | 0.2747 |
| *PyRapt* | Normalised crosscorrelation | 1.8122 |
| *PySWIPE* | Modified crosscorrelation | 14.6606 |

**Figure 4:** F0 spectra 0...20 Hz for the same signal with F0 estimators S0FT, YAAPT, Praat (cross-correlation), AMDF, PyRAPT, SWIPE (left column before right).

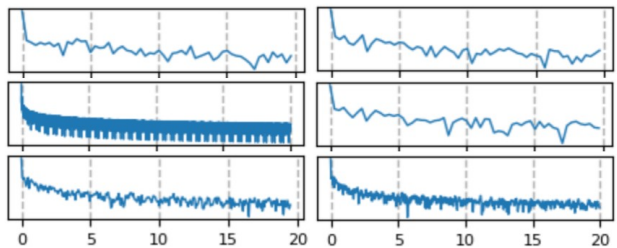

A bonus of the FES visualisation for critical phonetic pedagogy is the emergence of vastly different spectra obtained under the same conditions from the different and evidently only superficially similar F0 estimation algorithms, mainly due to their different windowing length and skip sizes, and on time vs. frequency domain estimation techniques (cf. the F0 spectra of F0 estimators in Figure 4).

Consequently, further calculation with the output values of the estimators must be based on an exact critical understanding of the algorithm properties and not simply based on a visual impression of plausibility or similarity.

## 6. SUMMARY, CONCLUSION, OUTLOOK

The novel *CRAFT* interactive online speech prosody teaching tool combines a coherent selected set of applications as an aid to a critical understanding of state of the art acoustic features of speech prosody. *CRAFT* is a tool for visualising concepts which are needed for understanding more advanced software and supplements acoustic phonetics teaching with more traditional media. *CRAFT* visualises tonal and rhythmic properties of speech, including polynomial models of pitch tendencies, AM and FM spectra and difference spectra, and illustrations of quantitative relations between AM and FM envelopes and between AM and FM spectra. The software source code is freely available in the GitHub standard repository.

The present contribution focuses solely on uses of visualisation in phonetic pedagogy. Research using *CRAFT* is currently in progress [13] on prosodic regularities in speech rhythm and fundamental frequency beyond syllable and foot levels, the 'slow rhythms' of utterances and discourse.

## 7. REFERENCES

[1] d'Allessandro, C., Mertens, P. 1995. Automatic pitch contour stylization using a model of tonal perception. *Computer Speech & Language* 9 (3), 257-288.
[2] Arjmandi, M. K., L. C. Dilley, M. Lehet. 2018. A comprehensive framework for F0 estimation and sampling in modeling prosodic variation in infant-directed speech. *Proc. 6th International Symposium on Tonal Aspects of Language*. Berlin, Germany.
[3] Barbosa, P. A. 2002. Explaining cross-linguistic rhythmic variability via a coupled-oscillator model of rhythm production. *Speech Prosody 2002,* 163-166.
[4] Boersma, P. 2001. Praat, a system for doing phonetics by computer. *Glot International* 5:9/10, 341-345.
[5] Camacho, A. 2007. SWIPE: A Sawtooth Waveform Inspired Pitch Estimator for speech and music. Ph.D. thesis, University of Florida.
[6] Cummins, F., Port, R. 1998. Rhythmic constraints on stress timing in English. *Journal of Phonetics* (1998) 26, pp. 145–171.
[7] De Cheveigné, A., & Kawahara, H. 2002. YIN, a fundamental frequency estimator for speech and music. *J.Ac.Soc.Am.* 111 (4), 1917-1930.
[8] De Looze, C., Hirst, D. 2010. Integrating changes of register into automatic intonation analysis. *Speech Prosody 4*.
[9] Fujisaki, H., Hirose, K. 1984. Analysis of voice fundamental frequency contours for declarative sentences of Japanese. *J.Ac.Soc.Japan (E)* (4), pp. 233–242.
[10] Garg, D. SWIPE pitch estimator. https://github.com/dishagarg/SWIPE [*PySWIPE*]
[11] Gaspari, D. 2016. Mandarin Tone Trainer. Masters thesis, Harvard Extension School. https://github.com/dgaspari/pyrapt

[12] Gibbon, D. 2018. The Future of Prosody: It's about Time. Keynote, *Proc. Speech Prosody 9*. https://www.isca-speech.org/archive/SpeechProsody_2018/pdfs/_Inv-1.pdf
[13] Gibbon, D. 2019. Rhythm Zone Theory: Speech Rhythms are Physical after all. https://arxiv.org/abs/1902.01267
[14] Guyot, P. Fast Python implementation of the Yin algorithm. https://github.com/patriceguyot/Yin/
[15] Hermansky, Hynek. 2010. History of modulation spectrum in ASR. *Proc. ICASSP 2010*.
[16] Hess, W. 1983. *Pitch Determination of Speech Signals: Algorithms and Devices*. Berlin: Springer.
[17] Hirst, D. and R. Espesser. 1993. Automatic modelling of fundamental frequency using a quadratic spline function. Travaux de l'Institut de Phonétique d'Aix 15. 75-83.
[18] Inden, B., Malisz, Z., Wagner, P., Wachsmuth, I. 2012. Rapid entrainment to spontaneous speech: A comparison of oscillator models. In Miyake, N., Peebles, D., Cooper, R. P., ds., *Proceedings of the 34th Annual Conference of the Cognitive Science Society*. Austin, TX: Cognitive Science Society.
[19] Jassem, Wiktor, Henryk Kubzdela, Piotr Domagała. 1983. *Segmentacja Sygnału Mowy na podstawie zmian Rozkładu Energii w Widmie [Speech Signal Segmentation based on changes in Energy Distribution in the Spectrum]*. Warsaw: Polska Akademia Nauk.
[20] Jouvet, D., Laprie, Y. 2017. Performance analysis of several pitch detection algorithms on simulated and real noisy speech data. *25th European Signal Processing Conference*.
[21] Kentner, G. 2017. Rhythmic parsing. *Linguistic Review, 34* (1), 123-155.
[22] Kim, J. W., Salamon, J., Li, P., Bello, J. P. 2018. CREPE: A convolutional representation for pitch estimation. ICASSP 2018.
[23] Klessa, K. 2016. Annotation Pro. Enhancing analyses of linguistic and paralinguistic features in speech. Wydział Neofilologii UAM, Poznań.
[24] Krause, M. 1984. Recent developments in speech signal pitch extraction. In: Gibbon, D., Richter, H., eds. *Intonation, Accent and Rhythm*. Berlin: De Gruyter, 243-252.
[25] Kuczmarski, T., Duran, D., Kordek, N., Bruni, J. 2013. Second-degree polynomial approximation of Mandarin Chinese lexical tone pitch contours -- a preliminary evaluation. In Wagner, P. (Hrsg.): *Elektronische Sprachsignalverarbeitung 24*. TUDpress, 218-222.
[26] Martin, P. 1996. WinPitch: un logiciel d'analyse temps réel de la fréquence fondamentale fonctionnant sous Windows, *Actes des XXIV Journées d'Étude sur la Parole, Avignon*. 224-227.
[27] Mertens, P. 2004. The Prosogram: semi-automatic transcription of prosody based on a tonal perception model. *Speech Prosody 2*.
[28] Pierrehumbert, J. B. The Phonology and Phonetics of English Intonation. Ph.D. thesis, MIT, 1980.
[29] Rabiner L. R., Cheng, M. J., Rosenberg, A. E. McGonegal, C. A. 1976. A comparative performance study of several pitch detection algorithms. *IEEE Transactions on Acoustics, Speech, and Signal processing*. ASSP-24 5.
[30] Reichel, U. 2018. CoPaSul Manual: Contour-based, parametric, and superpositional intonation stylization. V04. https://arxiv.org/abs/1612.04765
[31] Schmitt, B. J. B. AMFM_decompy. [PyYAAPT] https://github.com/bjbschmitt/AMFM_decompy
[32] Sjölander, K., Beskow, J. 2000. Wavesurfer – an open source speech tool. *Proc. Interspeech*, 464-467. http://www.speech.kth.se/wavesurfer/
[33] Talkin, D. 1995. A Robust Algorithm for Pitch Tracking (RAPT). In Kleijn, W. B., Palatal, K. K. eds. *Speech Coding and Synthesis*. Elsevier Science B.V., 497-518.
[34] Talkin, D. 2014. Reaper: Robust Epoch And Pitch EstimatoR. https://github.com/google/REAPER
[35] 't Hart, J., Collier, R., Cohen, A. 1990. *A perceptual study of intonation. An experimental phonetic approach to speech melody.* Cambridge: Cambridge University Press.
[36] Patavee, C.. Somchai, J., Visarut, A., Ekkarit, M. 2001. F0 feature extraction by polynomial regression function for monosyllabic Thai tone recognition, In: *Proc. Eurospeech, 2001*.
[37] Tilsen S., Johnson, K. Low-frequency Fourier analysis of speech rhythm. *J.Ac.Soc.Am*. 2008; 124 (2):EL34–EL39. [PubMed: 18681499]
[38] Tilsen, S., Arvaniti, A. Speech rhythm analysis with decomposition of the amplitude envelope: Characterizing rhythmic patterns within and across languages. *J.Ac.Soc.Am.* 134, 628 .2013.
[39] Todd, N. P. M., Brown, G. J. 1994. A computational model of prosody perception. *ICSLP 94*, 127-130.
[40] Wong, P-F., Siu, M-H. 2002. Integration of tone related features for Mandarin speech recognition by a one-pass search algorithm. *Proc. 6th International Conference on Signal Processing*, Beijing, China.
[41] Xu, Y. 2013. ProsodyPro – a tool for large-scale systematic prosody analysis. *Tools and Resources for the Analysis of Speech Prosody (TRASP 2013)*, Aix-en-Provence, France. 7-10.
[42] Zahorian, S. A., Hu, H. A spectral/temporal method for robust fundamental frequency tracking. *J.Ac.Soc.Am.* 123 (6), June 2008. [YAAPT]
[43] Zhao, Y. 2017. Prosody-based automated evaluation of degree of nativeness for L2 English learners: In the case of Tianjin Dialect speakers. Bachelors thesis, Tianjin: Tianjin University.